\documentclass[lettersize,journal]{IEEEtran}
\usepackage{amsmath,amsfonts}
\usepackage{array}
\usepackage[caption=false,font=normalsize,labelfont=sf,textfont=sf]{subfig}
\usepackage{textcomp}
\usepackage{cuted}
\usepackage{dblfloatfix}
\usepackage{amsmath}
\newcommand{\WideEqSep}{\par\vspace{0.4ex}\hrule\vspace{0.6ex}\par}
\usepackage{url}
\usepackage{verbatim}
\usepackage{graphicx}
\usepackage{setspace}
\usepackage{algorithm}
\usepackage{algpseudocode}
\usepackage{amsmath, amssymb}
\algrenewcommand\algorithmicrequire{\textbf{Input:}}
\algrenewcommand\algorithmicensure{\textbf{Output:}}
\algnewcommand{\LineComment}[1]{\State \(\triangleright\) #1}
\usepackage{caption}

\newtheorem{theorem}{\it Theorem}

\usepackage[caption=false,font=normalsize,labelfont=sf,textfont=sf]{subfig}
\def\BibTeX{{\rm B\kern-.05em{\sc i\kern-.025em b}\kern-.08em
		T\kern-.1667em\lower.7ex\hbox{E}\kern-.125emX}}
\usepackage{balance}
\begin{document}
	\title{\fontsize{18pt}{24pt}\selectfont{Multi-User Covert Communications via Intelligent Spectrum Control}}
	\author{Yujie Ling,~\IEEEmembership{Graduate Student Member,~IEEE}, Zan Li,~\IEEEmembership{Fellow,~IEEE}, Lei Guan,~\IEEEmembership{Member,~IEEE},\\
		\vspace{0.1em} Zheng Zhang,~\IEEEmembership{Member,~IEEE}, Dusit Niyato,~\IEEEmembership{Fellow,~IEEE}\\    \vspace{-1.2em}
			
			\thanks{Yujie Ling, Zan Li, Lei Guan, and Zheng Zhang are with the State Key Laboratory of Integrated Services Networks, Xidian University, Xi'an 710071, China (e-mail: yujieling@stu.xidian.edu.cn; zanli@xidian.edu.cn; lguan@xidian.edu.cn; zhangzheng@xidian.edu.cn).
				
				Dusit Niyato is with College of Computing and Data Science, Nanyang Technological University, Singapore (e-mail: dniyato@ntu.edu.sg).
				
		}}

		{}
		
		\maketitle
		
		\begin{abstract}
			This paper investigates the performance of multi-user covert communications over a fixed bandwidth in a multi-cell scenario with both eavesdroppers and malicious jammers. We propose an intelligent spectrum control (ISC) scheme that combines high-accuracy spectrum sensing with AI-assisted real-time decision-making to generate time-frequency dynamic occupation patterns for multiple legitimate users. The scheme can proactively avoid external interference and intra-system co-channel collisions, thereby improving covertness and reliability. Within this framework, we derive closed-form expressions for the detection error probability (DEP) of the eavesdropper and the reliable transmission probability (RTP) of legitimate users under multi-user joint detection. We then analytically optimize the transmission power that can maximize the covert rate (CR), as well as the maximum number of users that can access the system covertly and concurrently under given covertness and reliability constraints. Simulation results confirm the tight match between the analytical and Monte Carlo curves, and show that the proposed scheme can achieve a higher DEP, a larger RTP, and a greater multi-user capacity than the benchmark scheme.
		\end{abstract}
		
		
		\vspace{-4pt}
		\section{Introduction}
		With the rapid development of cellular networks and the Internet of Things (IoT), wireless communication has become deeply embedded in many critical applications. Unlike wired links, wireless channels are inherently open and broadcast. Radio signals propagate in space in a non-directional manner, allowing potential adversaries to passively listen or actively monitor legitimate transmissions without being noticed. Once a transmission is identified and possibly localized, legitimate nodes may still be exposed to jamming and physical attacks. In this context, covert communication has emerged as an important research direction~\cite{b1,b2}. Its main objective is to hide the presence of communication, so that a warden cannot distinguish while service requirements are still satisfied.
		
		In recent years, extensive efforts have been devoted to modeling and analyzing covert communications. The authors of \cite{b3} considered a full-duplex relay system and formulated an optimization framework that jointly designs power control and relay strategies. Based on this, they derived closed-form expressions for the minimum detection error probability (DEP) and maximum achievable covert rate (CR). The authors of \cite{b4} investigated an integrated sensing and communication system with dual-functional artificial noise (AN). By jointly designing the data and AN beams, they maximized the worst-case CR and demonstrated the potential of AN to enhance covertness performance. The authors of \cite{b5} studied a downlink multi-antenna non-orthogonal multiple access (NOMA) system with channel uncertainty. According to the minimum DEP and optimal detection threshold, they then maximized the covert throughput via power allocation and antenna selection. The authors of \cite{b6} proposed a phase-perturbed on-off keying (OOK) modulation scheme, where carefully designed phase offsets are superimposed on the OOK symbols. The legitimate receiver can still decode reliably, while warden cannot distinguish symbol distributions. The authors of \cite{b7} focused on relay-assisted IoT scenarios and developed a finite-blocklength covert communication framework based on optimal likelihood-ratio detection and KL-divergence constraints. However, most of them focus on a single link and usually assume that only one legitimate user attempts to access the channel at a time.
		
		In emerging multi-cell multi-user systems, a large number of base stations (BSs) and user terminals share fragmented spectrum resources under strong interference and asynchronous operation. In such environments, inter-cell spectrum reuse and clock misalignment lead to strong co-channel and adjacent-channel interference. This raises a fundamental question: \textit{Under the given constraints of covertness and reliability, how can multiple users simultaneously and covertly access the communication system without creating harmful mutual interference?} Different from the conventional single-link models, covert performance and communication quality in a multi-user environment not only depends on channel statistics and the detection method of the eavesdropper, but is also strongly influenced by the number of active users, their time-frequency occupation patterns, and the underlying scheduling scheme.
		
		Motivated by the above observations, we propose an intelligent spectrum control (ISC) scheme to enable covert access for multiple users. In each transmission period, the scheme employs intelligent decision-making modules to generate a dynamic frequency slot occupation pattern for all legitimate users, and it is driven by refined spectrum sensing methods so that malicious jammers and intra-system co-channel conflicts can be proactively avoided. Moreover, we analyze and derive the DEP of the eavesdropper under multi-slot joint detection, and the reliable transmission probability (RTP) of legitimate users by employing the proposed scheme. Building on these analytical results, we further analyze and determine the maximum CR with optimal transmission power, as well as the maximum number of users who can access the system covertly and concurrently under covertness and reliability constraints. Ultimately, we present the simulation results to prove the effectiveness of the proposed ISC transmission scheme.

		\section{System Model}
\begin{figure}[t!]
			\centerline{\includegraphics[width=1\linewidth]{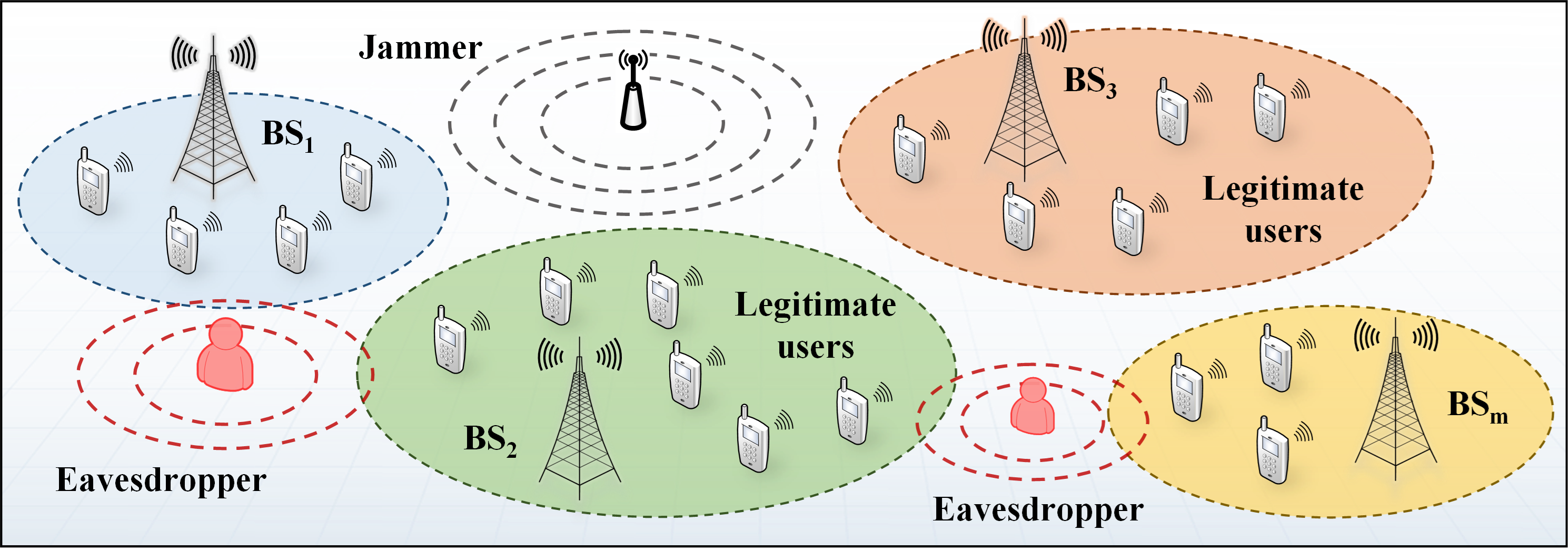}}
			\setlength{\belowcaptionskip}{-0.5cm}   
			\caption{Illustration of the multi-user covert communication scenario.}
			\label{f1}
		\end{figure}

		We consider a multi-cell, multi-user communication scenario shown in Fig.~\ref{f1}. The network consists of multiple BSs whose coverage areas do not overlap, and each BS serves several legitimate users within its communication cell with a fixed transmit power $p_b$. The set of all BSs is denoted by $\mathcal{N_B} = \left\{ {{\rm{B}}{{\rm{S}}_1},{\rm{B}}{{\rm{S}}_2}, \ldots ,{\rm{B}}{{\rm{S}}_m}} \right\}$, where $m$ is the total number of BSs in this scenario. To enable covert communications under multi-user coexistence, all terrestrial BSs are equipped with uniform linear array (ULA) antennas, which provide flexible spatial-domain control of the user signals. Our main interest lies in the covertness and quality of information transmission over the downlink \cite{b8,b9,b10}. During one transmission period, each legitimate user remains associated with its serving BS, which delivers confidential messages to that user.
		
		To elaborate, we assume that all links experience independent flat Rayleigh fading, so that the corresponding channel power gains follow exponential distributions. The set of legitimate users is denoted by $\mathcal{N_U} = \left\{ {{{\rm{U}}_1},{{\rm{U}}_2}, \ldots,{{\rm{U}}_k}} \right\}$, where $k$ is the total number of legitimate users engaged in covert communication. Because the BSs may operate under different clocks due to their locations and traffic demands, they are not synchronized, which leads to co-channel interference among cells. In addition, the spectrum used by this system is a publicly shared resource, and unknown external devices may also occupy the same frequency band, thereby introducing uncontrolled interference. There may also exist some unfriendly jammers in the area, which deliberately emit interference signals and thus degrade the quality of legitimate links.

		More critically, as shown in Fig.~\ref{f1}, the eavesdropper randomly located within the network coverage employs energy detectors to sense the transmissions of legitimate users. They attempt to decode the confidential messages, and their behavior poses the dominant threat to the overall covert communication performance. In the following analysis, we focus on a single passive eavesdropper located at the worst-case position.
		
		\vspace{-4pt}
		\section{ISC Scheme}
		In order to coordinate covert communication among multiple users, we propose an ISC-assisted transmission scheme. Assume that legitimate users transmit data over a total bandwidth $\mathcal{W}$. This total bandwidth is evenly divided into $q$ non-overlapping frequency slots, which forms an available frequency slots set $\mathcal{F} = \left\{ {{f_1},{f_2}, \ldots ,{f_q}} \right\}$. During one transmission period, the communication timeline is segmented into $p$ discrete time slots $\mathcal{T} = \left\{ {{t_1},{t_2}, \ldots ,{t_p}} \right\}$, and each legitimate user needs to occupy ${{L}}\left({{L}} \le p\right)$ consecutive time slots to deliver one data block. In each time slot, a legitimate user can occupy at most one frequency slot for downlink transmission.
		
\begin{figure}[t!]
			\centerline{\includegraphics[width=1\linewidth]{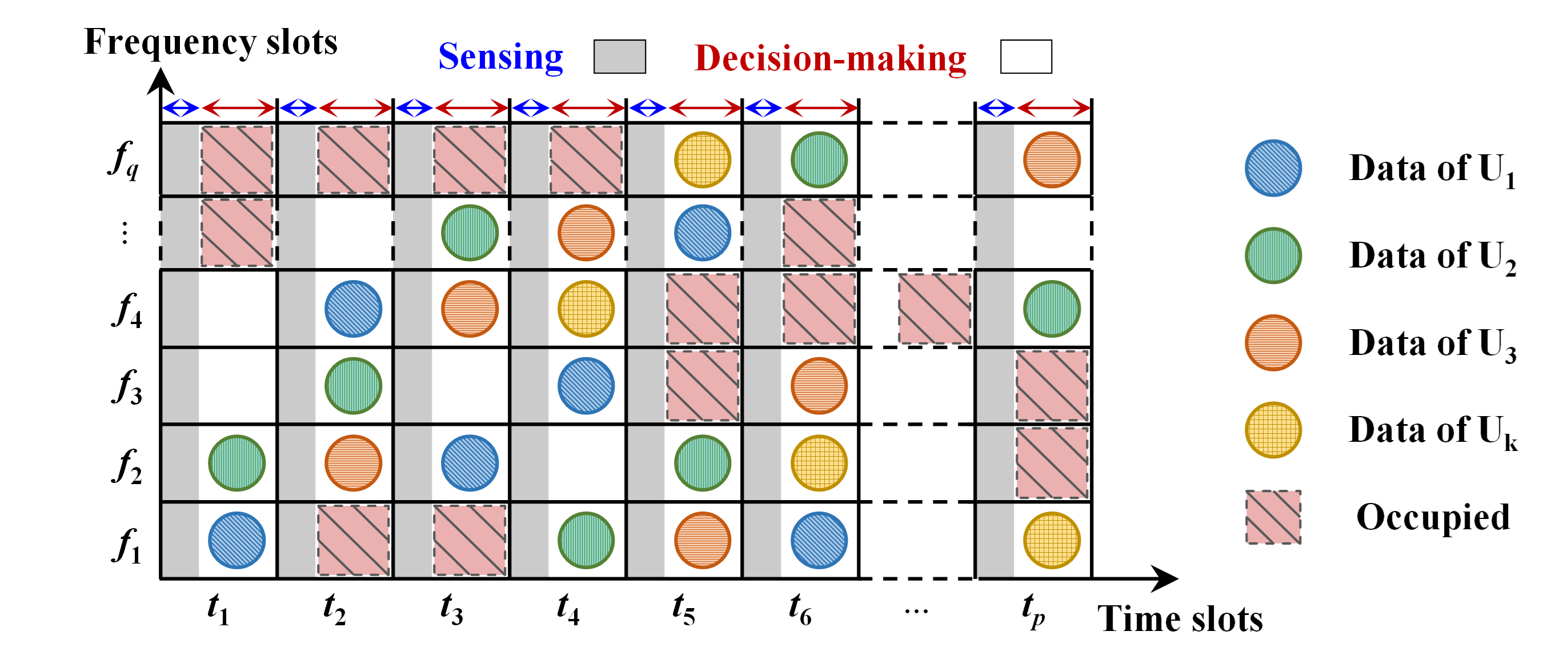}}
			\setlength{\abovecaptionskip}{-0.05cm}   
			\setlength{\belowcaptionskip}{-0.5cm}   
			\caption{Illustration of the proposed ISC scheme.}
			\label{f2}
		\end{figure}

		To enhance spectrum efficiency and suppress interference, we divide each time slot into a sensing phase and a decision-making phase. The sensing phase is short and relies on high-accuracy spectrum sensing techniques, such as convolutional neural network (CNN)-based and support vector machine (SVM)-based sensing methods~\cite{b11}, to detect the occupation state of each frequency slot. In decision-making phase, each BS is treated as an intelligent agent and trained by a deep double Q-network (DDQN) to generate a control matrix that guides all users in frequency slot occupation~\cite{b12,b13}. The frequency slot occupation status and channel quality indicators obtained in sensing phase form the state input of these agents. From the control matrix, the BS can directly check whether a frequency slot is occupied by more than one user and such conflicting allocations are penalized as negative feedback. Good channel conditions are regarded as positive feedback. Based on these feedback, each agent outputs frequency slot occupation actions for its served users and ensures that each user changes to different frequency slots across different time slots. The structure of the proposed scheme is shown in Fig.~\ref{f2}.
		
		By applying this scheme, malicious jammers and co-channel interference among legitimate users controlled by different BSs can be identified in sensing phase and then bypassed in decision-making phase~\cite{b14}. Simultaneously, each BS coordinates the frequency slot occupation of its served users to suppress intra-cell co-channel interference. Hence, the proposed scheme provides all legitimate users with effective channels that are of high quality and low interference. Note that the sensing result may be imperfect. Thus, we treat it as a noisy observation and use historical observations to estimate the occupation status of each frequency slot. Furthermore, since legitimate users are allowed to control in both time and frequency dynamically, the proposed scheme also creates favorable conditions for covert communication. Even if an eavesdropper observes suspicious energy peaks or noisy signal fragments, it cannot reliably infer the complete frequency slot occupation pattern. Hence, it is difficult for the eavesdropper to reconstruct the complete confidential message.

		\vspace{-4pt}
		\section{Performance Analysis}
		In this section, we analyze the covert communication performance of the system in terms of the DEP of the eavesdropper and the RTP of the legitimate user, respectively.
		\vspace{-8pt}
		\subsection{Detection Error Probability}
		As discussed earlier, legitimate users can dynamically hop over the entire set of available frequency slots. In this case, if the eavesdropper is assumed to reside on a single frequency slot, the resulting DEP can only reflect whether there is an ongoing transmission on that frequency slot~\cite{b15}.
		
		To more effectively characterize the covert performance in a multi-user environment, we consider the joint detection of the data transmission for eavesdropper. Assume that eavesdropper has a wideband observation capability and can perform energy detection over all frequency slots. Based on the condition of independent and identically distributed energy across slots, the joint test statistic at the eavesdropper is given by
		\begin{equation}
			\small
			{T_e} = \sum\nolimits_{j = 1}^q {\sum\nolimits_{i = 1}^L {{{\left| {{y_{e,{f_j}}}\left[ i \right]} \right|}^2}} },
			\label{e1}
		\end{equation}
		where ${{y_{e,{f_j}}}\left[ i \right]}$ denotes the observation on frequency slot $f_j$ at the $i$-th sample and satisfies
		\begin{equation}
			\small
			{y_{e,{f_j}}}\left[ i \right] =  
			\begin{cases}
				{\sqrt {\frac{{m{p_b}}}{{qkL}}} {h_{e,{f_j}}}x\left[ i \right] + {n_0}\left[ i \right]},  & {{{\cal H}_1},}\\
				{{n_0}\left[ i \right]}, & {{{\cal H}_0},}\\
			\end{cases}
			\label{e2}
		\end{equation}
		with ${{\cal H}_1}$ denoting the hypothesis that exactly $k$ frequency slots ($k$ users) carry legitimate data while the remaining $q-k$ frequency slots contain noise and ${{\cal H}_0}$ denoting the hypothesis that all $q$ frequency slots contain noise only.
		
		Given the detection threshold ${{\gamma _e}}$ of the eavesdropper and the noise power $\sigma _0^2$, the false alarm (FA) probability is defined as the probability of deciding that a transmission is present when ${{\cal H}_0}$ holds, which can be obtain by Eq.~(\ref{e3}), and $Q\left( { \cdot, \cdot } \right)$ is the upper regularized incomplete Gamma function.
		
		\begin{figure*}[!b]   
			\WideEqSep
			
			\begin{equation}
				\small
				{P_{{\rm{FA}}}}\left( {{\gamma _e}} \right) = \Pr \left( {{T_e} > {\gamma _e}|{{{\cal H}_0}}} \right) = \Pr \left( {\sum\nolimits_{j = 1}^q {\sum\nolimits_{i = 1}^L {{{\left| {{n_0}\left[ i \right]} \right|}^2}} }  > {\gamma _e}} \right) = Q\left( {qL,\frac{{{\gamma _e}}}{{\sigma _0^2}}} \right).
				\label{e3}
			\end{equation}
			
			\WideEqSep
			
			\begin{equation}
				\small
				{P_{{\rm{MD}}}}\left( {{\gamma _e}} \right) = \Pr \left( {{T_e} < {\gamma _e}|{{{\cal H}_1}}} \right) = \Pr \left( {\underbrace {{{\sum\nolimits_{j = 1}^k {\sum\nolimits_{i = 1}^L {\left| {\sqrt {\frac{{m{p_b}}}{{qkL}}} {h_{e,{f_j}}}x\left[ i \right] + {n_0}\left[ i \right]} \right|} } }^2}}_X + \underbrace {{{\sum\nolimits_{j = 1}^{q - k} {\sum\nolimits_{i = 1}^L {\left| {{n_0}\left[ i \right]} \right|} } }^2}}_Y < {\gamma _e}} \right).
				\label{e4}
			\end{equation}
			
			\WideEqSep
			
			\begin{equation}
				\small
				\begin{aligned}
					{f_Z}\left( z \right) &= \int_0^z {{f_X}\left( x \right){f_Y}\left( {z - x} \right)dx}  = \frac{{{{\left( {\rho g + \sigma _0^2} \right)}^{ - kL}}{{\left( {\sigma _0^2} \right)}^{ - \left( {q - k} \right)L}}}}{{\Gamma \left( {kL} \right)\Gamma \left( {\left( {q - k} \right)L} \right)}}{e^{ - \frac{z}{{\sigma _0^2}}}}\int_0^1 {{t^{kL - 1}}{{\left( {1 - t} \right)}^{\left( {q - k} \right)L - 1}}{e^{\frac{{\rho g}}{{\sigma _0^2\left( {\rho g + \sigma _0^2} \right)}}tz}}zdt}\\
					& = \frac{1}{{\Gamma \left( {qL} \right){{\left( {\rho g + \sigma _0^2} \right)}^{kL}}{{\left( {\sigma _0^2} \right)}^{\left( {q - k} \right)L}}}}{z^{qL - 1}}{e^{ - \frac{z}{{\sigma _0^2}}}}{}_1{F_1}\left( {kL;qL;\frac{{\rho gz}}{{\sigma _0^2\left( {\rho g + \sigma _0^2} \right)}}} \right).
				\end{aligned}				
				\label{e5}
			\end{equation}
			
			\WideEqSep

			\begin{equation}
				\small
				\begin{aligned}
					\label{e6}
					{P_{{\rm{MD}}}}\left( {{\gamma _e}|g} \right) &= \frac{1}{{\Gamma \left( {qL} \right){{\left( {\rho g + \sigma _0^2} \right)}^{kL}}{{\left( {\sigma _0^2} \right)}^{\left( {q - k} \right)L}}}}\int_0^{{\gamma _e}} {{z^{qL - 1}}{e^{ - \frac{z}{{\sigma _0^2}}}}{}_1{F_1}\left( {kL;qL;\frac{{\rho gz}}{{\sigma _0^2\left( {\rho g + \sigma _0^2} \right)}}} \right)dz} \\
					&  = \frac{1}{{\Gamma \left( {qL} \right){{\left( {\rho g + \sigma _0^2} \right)}^{kL}}{{\left( {\sigma _0^2} \right)}^{\left( {q - k} \right)L}}}}\sum\limits_{n = 0}^{ + \infty } {\frac{{{{\left( {kL} \right)}_n}}}{{{{\left( {qL} \right)}_n}}}\frac{1}{{n!}}{{\left( {\frac{{\rho g}}{{\sigma _0^2\left( {\rho g + \sigma _0^2} \right)}}} \right)}^n}} \int_0^{{\gamma _e}} {{z^{n + qL - 1}}{e^{ - \frac{z}{{\sigma _0^2}}}}dz} \\
					& = {\left( {\frac{{\sigma _0^2}}{{\rho g + \sigma _0^2}}} \right)^{kL}}\sum\limits_{n = 0}^{ + \infty } {\frac{{{{\left( {kL} \right)}_n}}}{{n!}}} {\left( {\frac{{\rho g}}{{\rho g + \sigma _0^2}}} \right)^n}P\left( {qL + n,\frac{{{\gamma _e}}}{{\sigma _0^2}}} \right).
				\end{aligned}
			\end{equation}
			
			\WideEqSep

			\begin{equation}
				\small
				\begin{aligned}
					\label{e7}
					{P_{{\rm{MD}}}}\left( {{\gamma _e}} \right) &= {\int_0^{ + \infty } {\frac{1}{{{\omega _e}}}{e^{ - \frac{1}{{{\omega _e}}}g}}{{\left( {\frac{{\sigma _0^2}}{{\rho g + \sigma _0^2}}} \right)}^{kL}}\sum\limits_{n = 0}^{ + \infty } {\frac{{{{\left( {kL} \right)}_n}}}{{n!}}\left( {\frac{{\rho g}}{{\rho g + \sigma _0^2}}} \right)} } ^n}P\left( {qL + n,\frac{{{\gamma _e}}}{{\sigma _0^2}}} \right)dg \\
					&  = \sum\limits_{n = 0}^\infty  {\frac{{{{\left( {kL} \right)}_n}}}{{n!}}} P\left( {qL + n,\frac{{{\gamma _e}}}{{\sigma _0^2}}} \right)\int_0^{ + \infty } {\frac{1}{{{\omega _e}}}{e^{ - \frac{1}{{{\omega _e}}}g}}{{\left( {\frac{{\sigma _0^2}}{{\rho g + \sigma _0^2}}} \right)}^{kL}}{{\left( {\frac{{\rho g}}{{\rho g + \sigma _0^2}}} \right)}^n}} dg \\
					& = \frac{{\sigma _0^2}}{{\rho {\omega _e}}}\sum\limits_{n = 0}^\infty  {{{\left( {kL} \right)}_n}U\left( {n + 1,2 - kL,\frac{{\sigma _0^2}}{{\rho {\omega _e}}}} \right)} P\left( {qL + n,\frac{{{\gamma _e}}}{{\sigma _0^2}}} \right).
				\end{aligned}
			\end{equation}
			
			\WideEqSep
			
			\begin{equation}
				\small
				{P_e}\left( {{\gamma _e}} \right) = {P_{{\rm{FA}}}}\left( {{\gamma _e}} \right) + {P_{{\rm{MD}}}}\left( {{\gamma _e}} \right)=Q\left( {qL,\frac{{{\gamma _e}}}{{\sigma _0^2}}} \right)+\frac{{qkL\sigma _0^2}}{{m{p_b}{\omega _e}}}\sum\limits_{n = 0}^\infty  {{{\left( {kL} \right)}_n}U\left( {n + 1,2 - kL,\frac{{qkL\sigma _0^2}}{{m{p_b}{\omega _e}}}} \right)} P\left( {qL + n,\frac{{{\gamma _e}}}{{\sigma _0^2}}} \right).
				\label{e8}
			\end{equation}
			
		\end{figure*}
		
		Similarly, the miss detection (MD) probability is defined as the probability of declaring no transmission when ${{\cal H}_1}$ is true, which can be given by Eq.~(\ref{e4}). Let ${\rho  = \frac{{m{p_b}}}{{qkL}}}$ and define $g = {| {{h_{e,{f_j}}}}|^2}$, which follows an exponential distribution with parameter $\omega _e$. Based on a given value of $g$, the random variable $X$ follows a Gamma distribution with shape parameter $kL$ and scale related to ${\rho g + \sigma _0^2}$, while $Y$ follows a Gamma distribution with shape parameter $\left(q-k\right)L$ and scale ${\sigma _0^2}$. By convolving these two distributions, the probability density function of $Z=X+Y$ can be expressed as Eq.~(\ref{e5}), where ${}_1{F_1}\left( { \cdot ; \cdot ; \cdot } \right)$ is the Kummer confluent hypergeometric function and $\Gamma\left( { \cdot } \right)$ is the Gamma function. Based on the above expression, the conditional MD probability for a given $g$ can be obtained as Eq.~(\ref{e6}), where ${{{\left( {\cdot} \right)}_n}}$ denotes the Pochhammer symbol and $P\left( { \cdot , \cdot } \right)$ is the lower regularized incomplete Gamma function. Finally, by averaging over the exponential distribution of $g$, the MD probability can be written as Eq.~(\ref{e7}), and $U\left( { \cdot ; \cdot ; \cdot } \right) $ is the Tricomi confluent hypergeometric function. Therefore, the DEP of the eavesdropper is given by Eq.~(\ref{e8}).
		
		\vspace{-18pt}
		\subsection{Reliable Transmission Probability}
		\vspace{-4pt}
		The communication performance of the system is characterized by the RTP, which is defined as the probability that a legitimate user can successfully decode the received signal. When the proposed scheme is employed, the received signal at the legitimate user is mainly impaired by thermal noise. As a result, the signal-to-interference-plus-noise ratio (SINR) of the legitimate user can be written as
		\begin{equation}
			\small
			{\rm{SIN}}{{\rm{R}}_u} = \frac{{{{m}}{p_b}}}{{{{Lk}}\sigma _0^2}}\sum\nolimits_{i = 1}^L {\sum\nolimits_{j = 1}^q {\Pr \left( {{f_j}} \right){{\left| {{h_{u,f_j}}} \right|}^2}} },
			\label{e9}
		\end{equation}
		where ${\Pr \left( {{f_j}} \right)}$ denotes the probability that frequency slot $f_j$ is occupied, and $h={{{| {{h_{u,f_j}}}|}^2}}$ is the channel power gain between the BS and the user, which follows an exponential distribution with parameter $\omega _u$. Since the slot occupation probabilities satisfy ${\sum\nolimits_{j = 1}^q {\Pr \left( {{f_j}} \right)}}=1$, so when the decoding threshold of the legitimate user is $\gamma _u$, the RTP can be given by
		\begin{equation}
			\small
			{{P_u}\left( {{\gamma _u}} \right) = \Pr \left( {{\rm{SIN}}{{\rm{R}}_u} \ge {\gamma _u}} \right) = \exp \left( { - \frac{{{\gamma _u}{{k}}\sigma _0^2}}{{{{m}}{p_b}{\omega _u}}}} \right)}.
			\label{e10}
		\end{equation}

		\vspace{-4pt}
		\section{Optimal Power and Admission Capacity}
		Since we focus on covert communication in a multi-user scenario, in addition to the covert performance of the system, another key metric is the user capacity. Therefore, in this section, we first optimize the transmit power to maximize the achievable CR under reliability and covertness constraints. Based on this optimal power policy, we then investigate the maximum number of legitimate users that the system can accommodate under the imposed communication constraints.
		
		From the two performance metrics derived in Section IV, namely the DEP of eavesdropper ${P_e}\left( {{\gamma _e}} \right)$ and the RTP of legitimate user $P_u\left( {{\gamma _u}} \right)$, we assume that the system enforces the following constraints: ${P_e}\left( {{\gamma _e}} \right) \ge 1 - {\varepsilon _e}$, ${P_u}\left( {{\gamma _u}} \right) \ge 1 - {\varepsilon _u}$, where ${\varepsilon _e}$ and ${\varepsilon _u}$ are small positive constants close to zero. On this basis, we can present two theorems as follows.
		
		\begin{theorem}Under reliability and covertness constraints, the optimal transmit power of the BS is 
			\begin{equation}
				\small
				{p_b}^* = \sup \left\{ {{p_b} \in \left[ {{p_{\min }},{p_{\max }}} \right]:{P_e}\left( {{p_b}} \right) \ge 1 - {\varepsilon _e}} \right\}.
				\label{e11}
			\end{equation}
		\end{theorem}
		
		\vspace{2pt}
		\begin{IEEEproof}
			To obtain this, we consider the following power optimization problem that maximizes the CR:
			
			\vspace{-4pt}
			\begin{subequations}
				\setlength{\abovedisplayskip}{0.01pt}
				\small
				\setlength{\jot}{2pt}
				\renewcommand{\theequation}{\theparentequation{\alph{equation}}}
				\begin{align}
					{\mathop {\max }\limits_{{p_b}} } \quad &{R{\left( {{p_b}} \right)} = \mathbb{E}_h\left[{{{\log }_2}\left( {1 + {\rm{SIN}}{{\rm{R}}_u}} \right)}\right]  } \label{e12a}\\
					{\rm{s}}{\rm{.t}}{\rm{.}}\quad\;&{P_e}{\left( {{p_b}} \right)} \ge 1 - {\varepsilon _e}, \label{e12b}\\
					&{P_u}{\left( {{p_b}} \right)} \ge 1 - {\varepsilon _u}, \label{e12c}\\
					&{p_{\min }} \le {p_b} \le {p_{\max }}.\label{e12d}
				\end{align}
			\end{subequations}
			
			From Eq.~(\ref{e12a}), it can be verified that $R{\left( {{p_b}} \right)}$ is a monotonically increasing function of $p_b$. Therefore, the optimum must lie at the upper boundary of the feasible power interval.
			
			By substituting Eq.~(\ref{e8}) into Eq.~(\ref{e12b}), the power upper bound ${p}_{\rm{up}}$ based on the covertness constraint is given by
			\begin{equation}
				{{p}_{\rm{up}}} \buildrel \Delta \over = \sup \left\{ {{p_b} \in \left[ {{p_{\min }},{p_{\max }}} \right]:{P_e}\left( {{p_b}} \right) \ge 1 - {\varepsilon _e}} \right\}.
				\label{e13}
			\end{equation}
			And combining Eq.~(\ref{e10}) with Eq.~(\ref{e12c}), the power lower bound ${p}_{\rm{low}}$ based on the reliability constraint is given by
			\begin{equation}
				{p}_{\rm{low}} \buildrel \Delta \over =  - \frac{{{\gamma _u}k\sigma _0^2}}{{m{\omega _u}\ln \left( {1 - {\varepsilon _u}} \right)}}.
				\label{e14}
			\end{equation}
			
			As long as ${p}_{\rm{up}} \ge {p}_{\rm{low}}$, the optimization problem has a unique solution, and we can obtain
			\begin{equation}
				{p_b}^* = \min \left\{ {{p_{\max }},{p}_{\rm{up}}} \right\}={p}_{\rm{up}}.
				\label{e15}
			\end{equation}
			The above conclusion can be drawn because we usually set the allowable range of power settings $\left[ {{p_{\min }},{p_{\max }}} \right]$ relatively wide, so that the default is ${p_{\max }} \ge {p}_{\rm{up}}$.
			
			At this optimal power, the maximum average achievable CR at the legitimate user can be given by
			\begin{equation}
				\begin{aligned}
					\small
					{R^*} &= \int_0^{ + \infty } {{{\log }_2}\left( {1 + \frac{{m{p_b}^*}}{{k\sigma _0^2}}h} \right)\frac{1}{{{\omega _u}}}{e^{ - \frac{h}{{{\omega _u}}}}}dh}\\  &= \frac{1}{{\ln 2}}\exp \left( {\frac{{k\sigma _0^2}}{{m{\omega _u}{p_b}^*}}} \right){E_1}\left( {\frac{{k\sigma _0^2}}{{m{\omega _u}{p_b}^*}}} \right),
					\label{e16}
				\end{aligned}
			\end{equation}		
			where ${E_1}\left(\cdot\right)$ is the exponential integral function.
		\end{IEEEproof}

		\vspace{2pt}
		\begin{theorem}Under reliability and covertness constraints, the maximum admissible number of legitimate users is 
			\begin{equation}
				\small
				{k^*} = \max \left\{ {k \in \left\{ {1, \ldots ,q} \right\}:{p}_{\rm{up}}\left( k \right) \ge {p}_{\rm{low}}\left( k \right)} \right\}.
				\label{e17}
			\end{equation}
		\end{theorem}
		
		\vspace{2pt}
		\begin{IEEEproof}
			To obtain the maximum number of legitimate users that can be supported by the system, we establish and consider the following optimization problem:
			
			\vspace{-3pt}
			\begin{subequations}
				\setlength{\abovedisplayskip}{-3pt}
				\small
				\setlength{\jot}{2pt}
				\renewcommand{\theequation}{\theparentequation{\alph{equation}}}
				\begin{align}
					\max \quad &{{ k}} \label{e18a}\\
					{\rm{s}}{\rm{.t}}{\rm{.}}\quad\;&{{P_e}\left( {k,{p_b^*}} \right)} \ge 1 - {\varepsilon _e}, \label{e18b}\\
					&{{P_u}\left( {k,{p_b^*}} \right)} \ge 1 - {\varepsilon _u}, \label{e18c}\\
					& q \ge k\ge1. \label{e18d}
				\end{align}
			\end{subequations}
			
			By inspecting the constraint structure of this optimization problem, we can see that it has the same form as Eq.~(12). In \textit{Theorem 1}, we first fix $k$ and optimize $p_b$, which leads to a power upper bound ${p}_{\rm{up}}$ from the covertness constraint and a power lower bound ${p}_{\rm{low}}$ from the reliability constraint. The optimal transmit power ${p_b}^*$ exists only when the feasible power interval is non-empty, that is, when ${p}_{\rm{up}} \ge {p}_{\rm{low}}$. In contrast, Eq.~(18) fixes ${p_b}^*$ and then works in the reverse direction to determine the maximum user number $k^*$. Therefore, we explicitly regard ${p}_{\rm{up}}\left(\cdot\right)$ and ${p}_{\rm{low}}\left(\cdot\right)$ as functions of $k$. According to \textit{Theorem 1}, for any given $k$, once ${p}_{\rm{up}}\left(k\right) \ge {p}_{\rm{low}}\left(k\right)$ holds, constraints Eq.~(\ref{e18b}) and Eq.~(\ref{e18c}) are automatically satisfied. This is because the covertness constraint specifies ${p}_{\rm{up}}\left(k\right)$, while the reliability constraint specifies ${p}_{\rm{low}}\left(k\right)$, and any element within their intersection satisfies both constraints at the same time. Hence, the maximum number of users is interpreted as the largest $k$ for which the feasible power interval remains non-empty, \textit{i.e.}, the largest $k$ satisfying ${p}_{\rm{up}}\left( k \right) \ge {p}_{\rm{low}}\left( k \right)$. By further incorporating Eq.~(\ref{e18d}), we can obtain the maximum number of legitimate users $k^*$ that can be accommodated.
		\end{IEEEproof}

		To this end, for a practical scenario with given performance requirements, the derived results can provide both the optimal power and maximum admission capacity. By properly controlling the number of active links with optimal power, the system can remain covert while improving the probability that multiple users successfully deliver their private data.

		\vspace{-2pt}
		\section{Numerical Results}
		We present the simulation results to evaluate the performance of the proposed ISC scheme in a multi-user scenario. In the simulations, we consider a multi-cell network with 4 BSs, where the total bandwidth is 500 MHz. The network also includes one intentional jammer and one passive eavesdropper. Further, we set $L=8$, $q=64$,$\gamma_e=2$ $\gamma_u=2$, $\omega _u=2$, $\omega _e=1$, and compare the performance of the proposed scheme with that of the AN-aided OFDM transmission scheme~\cite{b16}.
		
		\begin{figure}[t!]
			\centerline{\includegraphics[width=0.75\linewidth]{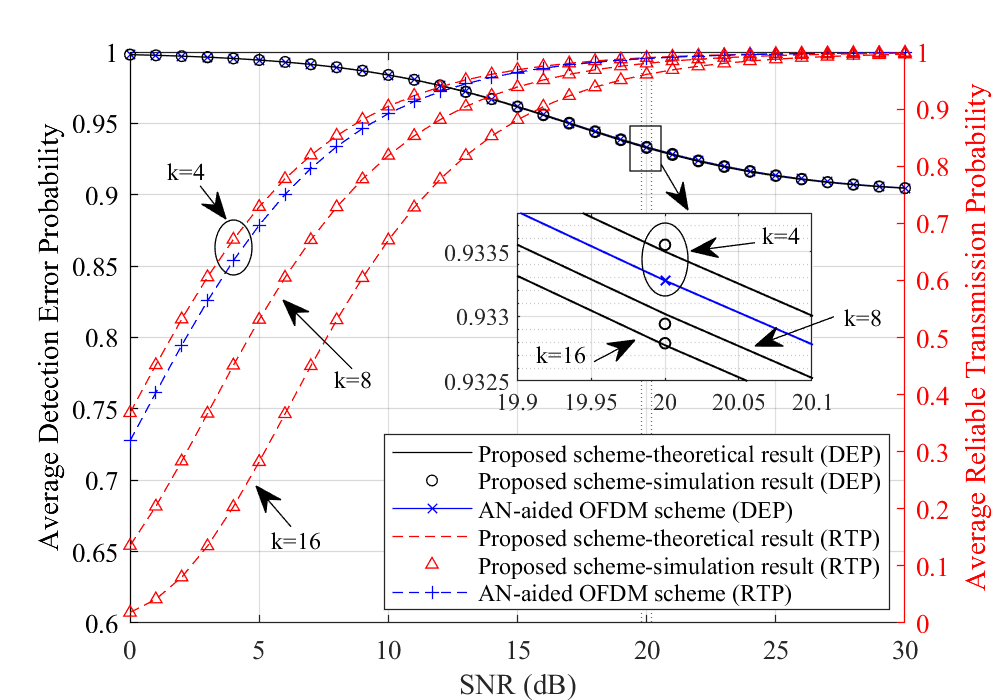}}
			\setlength{\belowcaptionskip}{-0.5cm}   
			\caption{Performance comparisons vs. different SNR.}
			\label{f3}
		\end{figure}
		
		Fig.~\ref{f3} shows the DEP and RTP of different schemes as SNR varies, when $k=4,8,16$. It can be observed from Fig.~\ref{f3} that the analytical results match the simulation points very well, which validates the correctness of our theoretical derivations. Fig.~\ref{f3} also shows that, for all schemes, the DEP decreases with increasing SNR, while the RTP increases as SNR grows. The reduction in DEP is due to the improvement of the observation quality at the eavesdropper. As the energy observation becomes higher, the eavesdropper can more easily distinguish active transmission from background noise, which enhances the detectability of the system and therefore reduces its covertness. In contrast, the growth in the RTP comes from the improved quality of the legitimate link, where the desired signal becomes stronger relative to noise, so that the receive SINR is more likely to exceed the decoding threshold and the events of decoding error or outage occur less frequently.
		
		From Fig.~\ref{f3}, we also find that, when the legitimate users number $k$ increases from 4 to 16, both the DEP and RTP decrease for all schemes. On the one hand, a larger number of active users implies that legitimate transmission signals occupy more frequency slots, so the transmit energy becomes stronger in the frequency domain. This makes it easier for eavesdropper to recognize the communication activity, and thus reduces the level of covertness. On the other hand, more users must share the same power and spectrum budget and introduce stronger co-channel interference and resource contention. As a result, the effective receive SINR of each legitimate link is reduced. Furthermore, compared with the benchmark scheme, the proposed scheme can achieve both a higher DEP and a larger RTP. Specifically, the AN-aided OFDM scheme reduces the SINR of eavesdropper by injecting AN. Too often, this scheme occupies a contiguous frequency band with a fixed set of subcarriers. In contrast, our proposed ISC scheme performs intelligent spectrum control over multiple frequency slots and follows a dynamic occupation pattern, which make it much harder for eavesdropper to intercept transmitted privacy information. At the same time, our scheme exploits advanced spectrum sensing methods to improve the frequency slot occupation decisions, which enables proactive avoidance of external interference and co-channel collisions, thereby improving the RTP.
		
		\begin{figure}[t!]
			\centerline{\includegraphics[width=0.75\linewidth]{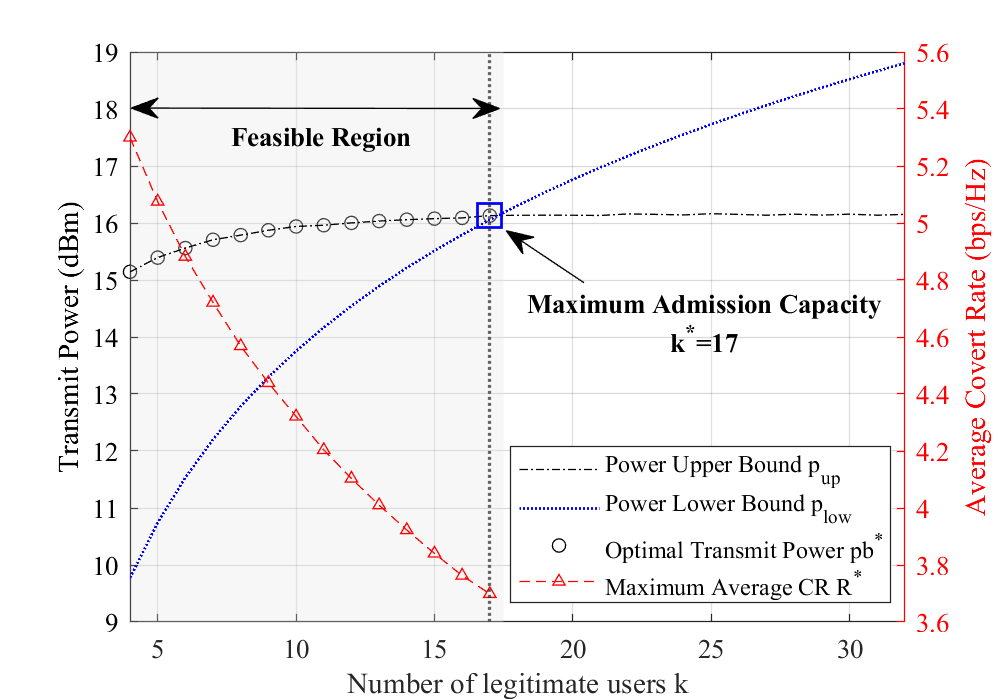}}
			\setlength{\belowcaptionskip}{-0.5cm}   
			\caption{Performance analyses vs. different user numbers.}
			\label{f4}
		\end{figure}
		
		Fig.~\ref{f4} shows, for different $k$, the optimal transmit power and the corresponding maximum CR achieved by the proposed scheme under joint covertness and reliability constraints. It can be observed that, as $k$ grows, both the upper and lower bounds of the feasible power region move to higher values. To satisfy the given reliability requirement, the BS has to increase its transmit power to compensate for the degradation of the link quality, which pushes the power lower bound ${p}_{\rm{low}}$ upwards as $k$ increases. Under the covertness constraint, the system can therefore tolerate a higher transmit power while still keeping the DEP above the prescribed level, which is reflected in the rise of the power upper bound ${p}_{\rm{up}}$ with respect to $k$. Fig.~\ref{f4} also shows that a feasible optimal transmit power and the associated maximum average CR exist only when the power upper bound is larger than the lower bound. Within this feasible region, both the covertness and reliability constraints can be satisfied, and the maximum admission capacity is determined by the largest $k$ for which the power region remains non-empty.

\begin{figure}[t!]
			\centerline{\includegraphics[width=0.75\linewidth]{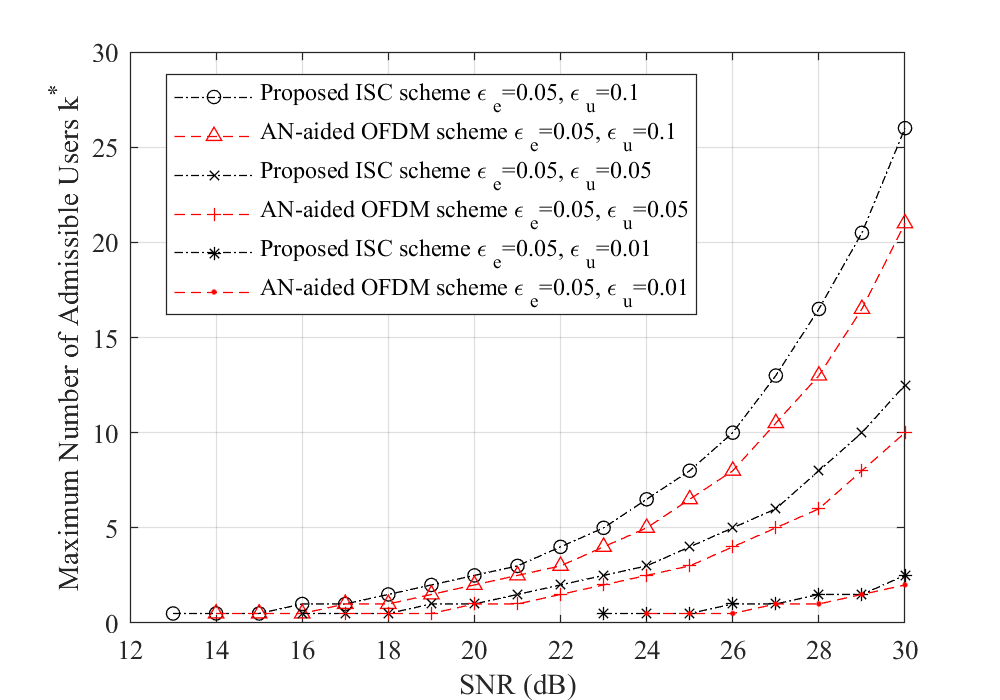}}
			\setlength{\belowcaptionskip}{-0.5cm}   
			\caption{Maximum legitimate user numbers vs. different SNR.}
			\label{f5}
		\end{figure}

		Fig.~\ref{f5} shows the maximum number of admissible legitimate users $k^*$ for different schemes as SNR varies. From Fig.~\ref{f5}, we can find that, as SNR increases, the maximum user number $k^*$ grows for all schemes. This is because the system obtains more additional SINR margin and can therefore admit more users to transmit simultaneously without degrading the individual decoding success probability. Fig.~\ref{f5} also shows that, maximum capacity $k^*$ increases as the reliability constraint becomes less stringent. The reason is that, when $\varepsilon _u$ increases, the system can sacrifice part of decoding reliability in exchange for a larger number of simultaneous users. Fig.~\ref{f5} also shows that the proposed scheme supports more legitimate users than the benchmark scheme. This is because the proposed design combines spectrum sensing with intelligent scheduling, which can proactively remove co-channel conflicts and external jammers without introducing additional power overhead. Consequently, it can satisfy both the covertness and reliability constraints while serving a larger set of legitimate users in parallel.
		
		\section{Conclusion and Future Work}
		In this paper, we have investigated a multi-user scenario and developed an ISC scheme to provide reliable and covert delivery of private information for multiple legitimate users. Within the proposed ISC framework, we have defined and analyzed the DEP and RTP. Based on these metrics, we have further derived the optimal power and the maximum number of legitimate users that can be supported. These results provide useful guidance for the design of covert access schemes in multi-user systems. For future work, we plan to incorporate cooperative jamming nodes into our framework, so as to enhance capacity while maintaining covertness and reliability.

	\end{document}